\documentclass[aps,preprint]{revtex4}
\usepackage[dvips]{graphics,graphicx}
\usepackage{amsmath}
\begin{document}

\title{The open Bose-Hubbard chain: Pseudoclassical approach}
\author{A.~A.~Bychek$^1$}
\author{P.~S.~Muraev$^2$}
\author{D.~N.~Maksimov$^{1,3}$}
\author{A.~R.~Kolovsky$^{1,2}$}
\affiliation{$^1$Kirensky Institute of Physics, 660036 Krasnoyarsk, Russia}
\affiliation{$^2$Siberian Federal University, 660041 Krasnoyarsk, Russia}
\affiliation{$^3$Siberian State Aerospace University, 660014 Krasnoyarsk, Russia}
\date{\today}
\begin{abstract}
We analyze stationary current of bosonic carriers in the Bose-Hubbard chain of length $L$ where the first and the last sites of the chain are attached to reservoirs of Bose particles acting as the particle source and sink, respectively. The analysis is curried out by using the pseudoclassical approach which reduces the original quantum problem to the classical problem for $L$ coupled nonlinear oscillators. It is shown that an increase of oscillator nonlinearity (which is determined by the strength of inter-particle interactions) results in a transition from the ballistic transport regime, where the stationary current is independent of the chain length,  to the diffusive regime, where the current is inverse proportional to $L$.  
\end{abstract}
\maketitle
 
\section{Introduction}

Transport phenomena with neutral atoms is an active area of research in the experimental  cold-atom physics. This includes Josephson oscillations of Bose atoms in a double-well potential \cite{Gati07,Este08}, atomic Bloch oscillations in quasi one-dimensional optical lattices \cite{Mein13,Fuji19}, refill dynamics in the presence of the induced losses \cite{Labo15,Labo16}, quantized current in the engineered transport channel  \cite{Bran12,Krin15,Lebr18} -- to mention a few of phenomena under scrutiny. From the theoretical viewpoint these experiments refer to (or even are aimed to realize) some paradigm models of the many-body physics that, on the one hand, are complex enough to capture the physics of the discussed phenomenon but, on the other hand, are simple enough to admit an analytical treatment. The present work analyzes one of these paradigm models, namely, the open or dissipative Bose-Hubbard (BH) chain \cite{Ivan13,Kord15a,Kord15b,112}. With respect to cold atoms this model describes atomic current between two reservoirs of Bose atoms which are connected by a deep optical lattice. More formally, the model assumes that the system dynamics is governed by the master equation for the reduced density matrix ${\cal R}(t)$ of the bosonic carriers,
\begin{equation}
\label{1}
\frac{d {\cal R}}{dt}=-i[\widehat{H},{\cal R}] + \widehat{{\cal L}}_1({\cal R}) +  \widehat{{\cal L}}_L({\cal R}) \;,
\end{equation}
where $\widehat{H}$ is the Hamiltonian of the closed (isolated) BH model,
\begin{equation}
\label{2}
\widehat{H}=\omega \sum_{l=1}^L  \hat{n}_l   
  -\frac{J}{2} \sum_{l=1}^{L-1} \left( \hat{a}^\dag_{l+1}\hat{a}_l +h.c.\right)
  +\frac{U}{2}\sum_{l=1}^L \hat{n}_l(\hat{n}_l-1) \;,
\end{equation}
and the relaxation operator ${\cal L}_1({\cal R}) $ acting on the first site of the chain  and the operator ${\cal L}_L({\cal R}) $ acting on the last site of the chain have the standard Lindblad form parametrized by the relaxation constant $\gamma_1$ ($\gamma_L$) and the particle density $\bar{n}_1$ ($\bar{n}_L$) of the respective reservoir,  see Eqs.~(\ref{a2}-\ref{a3}) below.  To be certain we assume  $\bar{n}_1>\bar{n}_L$, then the left reservoir acts as a particle source and the right reservoir as a particle sink. The goal is to find the stationary current $\tilde{j}$ of bosonic particles  in the chain as a function of the reservoir parameters and parameters of the Bose-Hubbard chain.  We also mention similar problems for the stationary current of the fermionic carries and heat transport in the spin chain \cite{Pros10,Pros14,Buca17}.

Without additional assumptions/approximations the formulated  problem can be solved only for vanishing interparticle interactions ($U=0$), where one obtains a relatively simple analytical expression for the stationary current \cite{112} . As expected, it appears to be proportional to the difference in the reservoir densities $\bar{n}_1-\bar{n}_L$ and is independent of the chain length $L$. The case of finite interactions $U\ne0$, however, remains a challenge. In particular, it is not clear whether the system is conducting or becomes insulating in the limit $L\rightarrow\infty$. In this work we analyze this problem by using the so-called pseudoclassical approach \cite{Mahm71,Moss06,Grae07,Trim08a,80,Zibo10,107,110} which is  justified for large occupation numbers of the chain sites and was proved to be successful in application to  a number of other problems like, for example,  dynamics of a Bose-Einstein condensates in the double-well potential \cite{Zibo10,110} or Bloch oscillations of interacting Bose atoms \cite{80}. The advantage of the pseudoclassical approach is that it provides a clear insight into the physics of the discussed phenomenon by mapping the quantum many-system into a classical (generally, multi-dimensional) system. Using this approach we show that an increase of the interaction constant from zero to a finite value $U\sim J/\bar{n}_1$ results in a transition from the ballistic transport regime, where the stationary current $\tilde{j}$ is independent of the chain length $L$,  to the diffusive regime, where $\tilde{j} \sim 1/L$.  

\section{Pseudoclassical approach}

To illustrate the main ideas of the pseudoclassical approach we revisit the problem of the damped anharmonic oscillator,
\begin{equation}
\label{a0}
\widehat{H}=\omega \hat{a}^\dagger \hat{a} + \frac{U}{2} (\hat{a}^\dagger \hat{a})^2  \;,
\end{equation}
where  we set the fundamental Planck constant to unity, i.e., $[\hat{a},\hat{a}^\dagger]=1$.

\subsection{Quantum analysis}
\begin{figure}[t]
\includegraphics[width=12.5cm,clip]{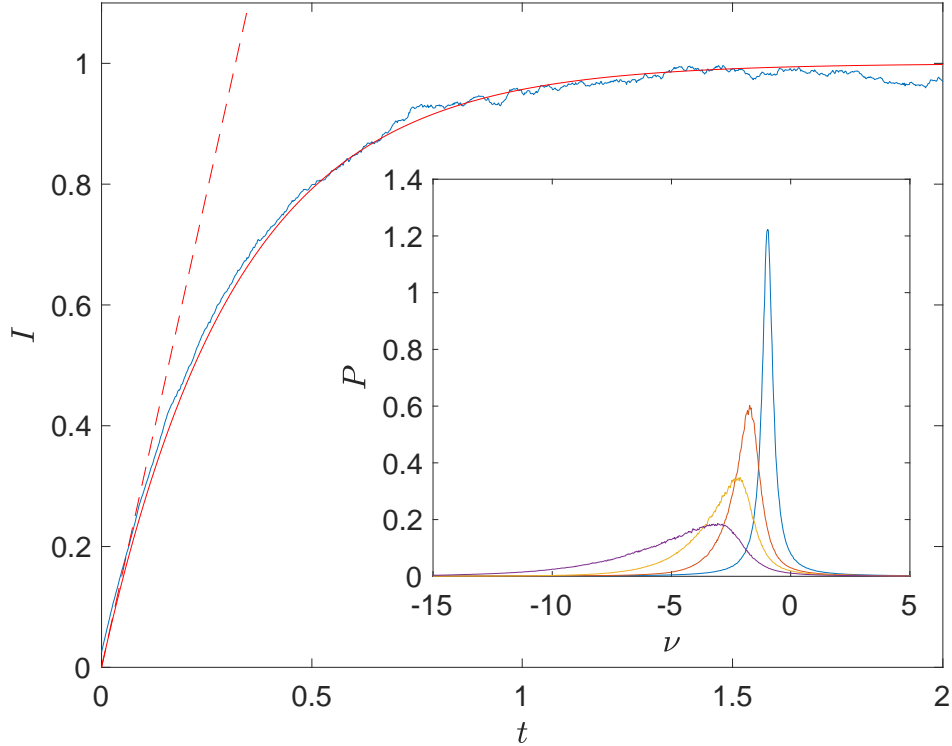}
\caption{Solid line: Dynamics of $I(t)=N(t)/\bar{n}$ according to Eq.~(\ref{a1}). Parameter are $\gamma=0.5$ and $\bar{n}=10$. The initial condition corresponds to the oscillator ground state, i.e., to the Fock state $| n\rangle$ with $n=0$. Dashed line: Solution of the master equation (\ref{a5}) for $D=0.5$ and $\gamma=0$. Broken line: Solution of the Langevin equation (\ref{b4}) averaged over 4000 realizations of the random process $\xi(t)$.  The inset shows the oscillator spectral density $P(\nu)$ for $\omega=1$ and different values of nonlinearity $g=0,0.5,1,2$, from top to bottom.}
\label{fig1}
\end{figure}

Commonly the dissipative dynamics of the system (\ref{a0})  is described by the following master equation 
\begin{equation}
\label{a1}
\frac{d {\cal R}}{dt}=-i[\widehat{H},{\cal R}] + \widehat{{\cal L}}_{gain}({\cal R}) + \widehat{{\cal L}}_{loss}({\cal R}) \;,
\end{equation}
where the Lindblad operators $\widehat{{\cal L}}_{loss}$,
\begin{equation}
\label{a2}
\widehat{{\cal L}}_{loss}({\cal R})=-\frac{\gamma(\bar{n}+1)}{2}
 (\hat{a}^\dagger \hat{a}{\cal R}-2\hat{a}{\cal R}\hat{a}^\dagger + {\cal R}\hat{a}^\dagger\hat{a}) \;,
\end{equation}
and $\widehat{{\cal L}}_{gain}$,
\begin{equation}
\label{a3}
\widehat{{\cal L}}_{gain}({\cal R})=-\frac{\gamma\bar{n}}{2}
 (\hat{a}\hat{a}^\dagger{\cal R}-2\hat{a}^\dagger{\cal R}\hat{a} + {\cal R}\hat{a}\hat{a}^\dagger) \;,
\end{equation}
take into account exchange of particles between the system and bosonic reservoir characterised by the parameter $\bar{n}$. It is easy to prove that the Lindblad operators  (\ref{a2}-\ref{a3}) insure relaxation of the oscillator to the steady state with the mean occupation number given by the parameter $\bar{n}$. As an example, the red solid line in   Fig.~\ref{fig1} shows the solution of Eq.~(\ref{a1}) for $\bar{n}=10$ and $\gamma=0.5$  where we choose the oscillator ground state as the initial condition. For this specific initial condition relaxation to the steady state has particularly simple time dependence
$N(t)={\rm Tr}[\hat{a}^\dagger \hat{a} {\cal R}(t)]=\bar{n}[1-\exp(-\gamma t)]$.

To get a deeper insight into the discussed relaxation process it is instructive to rewrite Eq.~(\ref{a1}) in the following form
\begin{equation}
\label{a5}
\frac{d {\cal R}}{dt}=-i[\widehat{H},{\cal R}] + \widehat{{\cal D}}({\cal R}) + \widehat{{\cal G}}({\cal R}) \;,
\end{equation}
where
\begin{equation}
\label{a6}
\widehat{{\cal D}}({\cal R})=-\frac{D\bar{n}}{2}\left( [\hat{a},[\hat{a}^\dagger,{\cal R}]] + [\hat{a}^\dagger,[\hat{a},{\cal R}]] \right)  \;, 
\end{equation}
and
\begin{equation}
\label{a7}
\widehat{{\cal G}}({\cal R})=-\frac{\gamma}{2}
 (\hat{a}\hat{a}^\dagger{\cal R}-2\hat{a}^\dagger{\cal R}\hat{a} + {\cal R}\hat{a}\hat{a}^\dagger) \;.
\end{equation}
Here, as it will be clear in a moment, the operator (\ref{a6}) is the diffusion term with the diffusion coefficient $D=\gamma$ and the operator (\ref{a7}) is the friction term with the friction coefficient $\gamma$. We  mention that in the present-day laboratory experiments with cold Bose atoms the diffusion and friction coefficients can be varied independently. Then, for $\gamma=0$ we have pure diffusion (see dashed line in Fig.~\ref{fig1}) while for $D=0$ we  have pure decay. In the general case an interplay between these processes leads to the steady state where
\begin{equation}
\label{a8}
\lim_{t\rightarrow\infty} N(t)\equiv \tilde{N}=\bar{n}\frac{D}{\gamma} \;.
\end{equation}

\subsection{Classical analysis}

Let us reproduce the result (\ref{a8}) by using the pseudoclassical approach. In this approach the operators $\tilde{a}=\hat{a}/\sqrt{\bar{n}}$ and $\tilde{a}^\dagger=\hat{a}^\dagger/\sqrt{\bar{n}}$, which commute to the effective Planck constant $\hbar'=1/\bar{n}$,
\begin{equation}
\label{b0a}
[\tilde{a},\tilde{a}^\dagger ]=\hbar' \;, \quad \hbar'=1/\bar{n} \ll 1\;,
\end{equation}
are identified with the pair of canonical variables $a$ and $a^*$,
\begin{equation}
\label{b0b}
a=(q+ip)/\sqrt{2} \;,\quad a^*=(q-ip)/\sqrt{2} \;.
\end{equation}
Applying the Wigner-Weyl transformation to the master equation (\ref{a5}),  and keeping only the zero-oder terms in the expansion over the effective Planck constant $\hbar'$ we obtain the following equation for the  Wigner function  $f=f(a,a^*;t)$ \cite{113},
\begin{equation}
\label{b1}
\frac{\partial f}{\partial t}=\{H,f\} + {\cal D}(f) + {\cal G}(f) \;,
\end{equation}
where the Hamiltonian evolution is captured by the Poisson brackets with the Hamiltonian
\begin{equation}
\label{b0}
H=\omega a^*a +\frac{g}{2} (a^*a)^2 \;, \quad g=U\bar{n} \;,
\end{equation}
the diffusion term is
\begin{equation}
\label{b2}
{\cal D}(f)= D\frac{\partial^2 f}{\partial a \partial a^*}  \;,
\end{equation}
and the friction term is given by
\begin{equation}
\label{b3}
{\cal G}(f)=\frac{\gamma}{2}\left( a\frac{\partial f}{\partial a} + 2f + a^*\frac{\partial f}{\partial a^*} \right) \;.
\end{equation}

The obtained Fokker-Planck Eq.~(\ref{b1}) has a simple analytical solution  if the initial distribution function is given by the two-dimensional Gaussian
\begin{equation}
\label{b5}
f(a,a^*; t=0)=\frac{1}{2\pi i \sigma^2}\exp\left(-\frac{|a|^2}{\sigma^2} \right) 
= \frac{1}{2\pi \sigma^2}\exp\left(-\frac{q^2+p^2}{2\sigma^2} \right) \;.
\end{equation}
Substituting this ansatz with $\sigma^2=\sigma^2(t)$ in Eq.~(\ref{b1}) we find that $\sigma^2(t)$ grows linearly if $\gamma=0$ or decays exponentially if $D=0$. Correspondently,  the mean oscillator action $I(t)$, 
\begin{equation}
\label{b6}
I(t)= \int \int |a|^2 f(a,a^*;t) {\rm d} a {\rm d} a^* \;,
\end{equation}
grows as $I(t)=I_0 + Dt$ if $\gamma=0$ or decays  as $I(t)=I_0\exp(-\gamma t)$  if $D=0$. In the general case the oscillator approaches the steady state  where 
\begin{equation}
\label{b7}
\lim_{t\rightarrow\infty} I(t)\equiv \tilde{I}=\frac{D}{\gamma} \;.
\end{equation}
Notice that  Eq.~(\ref{b7}) holds for arbitrary oscillator nonlinearity $g$.

\subsection{Stochastic method}

For the purpose of future reference we discuss the Monte-Carlo method of solving  Eq.~(\ref{b1}). It is easy to prove that the  Fokker-Planck Eq.~(\ref{b1}) is equivalent to the following Langevin equation
\begin{equation}
\label{b4}
i\dot{a}= \frac{\partial H}{\partial a^*} - i\frac{\gamma}{2} a + \sqrt{\frac{D}{2}} \xi(t)  \;,
\end{equation}
where $\xi(t)={\rm Re}[\xi(t)]+i {\rm Im}[\xi(t)]$ is the $\delta$-correlated noise with standard deviation equal to unity for both the real and imaginary parts.  Then the mean value of action Eq.~(\ref{b6}) can be calculated as the average over different realizations of the random process $\xi(t)$, namely,  $I(t)=\overline{|a(t) |^2}$. As an example the broken line in Fig.~\ref{fig1} shows the mean action which is obtained by using 4000 realizations of the random process. Increasing the number of realizations by one oder of magnitude makes the result practically indistinguishable from the exact solution.

Importantly, the described stochastic approach allows us to introduce the other important characteristic of the system -- the oscillator spectral density $P(\nu)$ \cite{Bixo71,book},
\begin{equation}
\label{b8}
P(\nu)=\overline{|a(\nu)|^2} \;, \quad a(\nu)=\lim_{T\rightarrow\infty} \frac{1}{T} \int_0^T a(t) e^{i\nu t} {\rm d}t \;.
\end{equation}
Here the limit $T\rightarrow\infty$ insures that $P(\nu)$ refers to the stationary regime. In practice this requires the evolution time $T$ to be larger than the relaxation time $T_\gamma \sim 1/\gamma$. If this condition is satisfied then the stationary action (\ref{b7}) is related to the spectral density as
\begin{equation}
\label{b9}
\tilde{I}=\int_{-\infty}^\infty P(\nu){\rm d} \nu \;.
\end{equation}
Notice that, unlike the stationary action $\tilde{I}$, the spectral density  $P(\nu)$ does depend on the oscillator nonlinearity. Different curves in the inset in Fig.~\ref{fig1} show spectral density of the oscillator (\ref{b0}) for $\omega=1$ and increasing value of $g$.  It is seen that with increase of $g$ the maximum of the spectral density shifts to the higher (negative) frequencies. A qualitative explanation for this shift is that for nonzero nonlinearity the oscillator eigenfrequency $\Omega=\omega+gI$ depends on the action $I$ which in the stationary regime is distributed according to the exponential law $f(I)=\sigma^{-2}\exp(-I/\sigma^2)$, $\sigma^2=D/\gamma$. On the quantitative level calculation of the spectral density requires sophisticated methods which involve continued fraction expansion \cite{Bixo71}. The only exclusion is the case $g=0$ where $P(\nu)$ can be calculated exactly by using formalism of the Green function \cite{book}, that gives 
\begin{equation}
\label{b5c}
P(\nu)=\frac{D}{2\pi}\frac{1}{(\nu+\omega)^2+(\gamma/2)^2} \;.
\end{equation}

\section{Chain of coupled oscillators}

Now we are prepared to address the current in the BH chain of length $L$ connecting two reservoirs.  Generalizing the results of the previous section for the system (\ref{2}), the equation for the classical distribution function $f=f({\bf a},{\bf a}^*;t)$, ${\bf a}=(a_1,\ldots,a_L)$, reads
\begin{equation}
\label{c0a}
\frac{\partial f}{\partial t}=\{H,f\} + \sum_{l=1,L}{\cal D}^{(l)}(f) +  \sum_{l=1,L}{\cal G}^{(l)}(f)  \;,
\end{equation}
where
\begin{equation}
\label{c0b}
{\cal D}^{(l)}(f)= D_l\frac{\partial^2 f}{\partial a_l \partial a_l^*}  \;,
\end{equation}
and  
\begin{equation}
\label{c0c}
{\cal G}^{(l)}(f)=\frac{\gamma_l}{2}\left( a_l\frac{\partial f}{\partial a_l} + 2f+ a_l^*\frac{\partial f}{\partial a_l^*} \right) \;.
\end{equation}
Thus, in the pseudo-classical approach the original quantum problem corresponds to the system of $L$ coupled nonlinear oscillators,
\begin{equation}
\label{c1}
H= \sum_{l=1}^L H_l  - \frac{J}{2}\left(\sum_l a^*_{l+1} a_l + c.c.\right) \;, \quad
H_l=\omega I_l +\frac{g}{2}I_l^2  \;,
\end{equation}
where the first and the last oscillators have non-zero friction coefficient $\gamma_1$ and $\gamma_L$ and are driven by stochastic forces with the amplitude $\sqrt{D_1/2}$ and $\sqrt{D_L/2}$, respectively,
\begin{equation}
\label{c2}
i\dot{a_l}= \frac{\partial H}{\partial a_l^*} - i\frac{\gamma_l}{2}(\delta_{1,l}+\delta_{L,l}) a_l 
+ \sqrt{\frac{D_l}{2}}(\delta_{1,l} + \delta_{L,l}) \xi_l(t) \;.
\end{equation}
%

\subsection{Single-particle density matrix}

To calculate the current of the bosonic carriers in the open BH chain it suffices to know the single-particle density matrix (SPDM) whose elements are defined as
\begin{equation}
\label{c3c}
\rho_{l,m}(t)={\rm Tr}[\hat{a}_l^\dagger\hat{a}_m {\cal R}(t)] \;,\quad 1\le l,m \le L \;.
\end{equation}
In the pseudoclassical approach Eq.~(\ref{c3c}) obviously takes the form 
\begin{equation}
\label{c3a}
\rho_{l,m}(t)= \int a_l^* a_m  f({\bf a},{\bf a}^*;t) {\rm d} {\bf a} {\rm d} {\bf a}^* \;,
\end{equation}
if we use the formalism of the Fokker-Planck equation (\ref{c0a}) or 
\begin{equation}
\label{c3b}
\rho_{l,m}(t)=\overline{a_l^*(t) a_m(t)} 
\end{equation}
if we use the Langiven equation (\ref{c2}). Knowing SPDM the mean current $j(t)$ is then found as
\begin{equation}
\label{c6}
j(t)={\rm Tr}[\hat{j} \hat{\rho}(t)]=J \sum_l {\rm Im}[\rho_{l,l+1}(t)] \;.
\end{equation}

Remarkably, if $g=0$ the matrix (\ref{c3a}) can be found analytically. Taking time derivative from both sides of  Eq.~(\ref{c3a}) and substituting there Eq.~(\ref{c0a}) we obtain after integrating by parts the following system of equations on the matrix elements,
\begin{eqnarray}
\nonumber
\frac{{\rm d}}{{\rm d} t} \rho_{l,m}=i\frac{J}{2}(\rho_{l,m+1} + \rho_{l,m-1} -\rho_{l+1,m} -\rho_{l-1,m}) 
-\sum_{j=1,L}\frac{\gamma_j}{2}(\delta_{l,j}+\delta_{m,j})\rho_{l,m} \\
\label{c3}
+ \sum_{j=1,L} D_j \delta_{l,j}\delta_{m,j} \;.
\end{eqnarray}
This equation coincides with that  obtained quantum-mechanically in Ref.~\cite{112}. Thus, in the case of vanishing inter-particle interactions the pseudoclassical approach is {\em exact}. The stationary solution of Eq.~(\ref{c3}) is the three diagonal matrix with pure imaginary off-diagonal elements which determine the stationary current $\tilde{j}$. In what follows we shall consider the case $\gamma_1=\gamma_L\equiv\gamma$ where analytical expression for the stationary current is particularly  simple, 
\begin{equation}
\label{c4}
\tilde{j}=\frac{J^2\gamma}{J^2+\gamma^2}\frac{D_1-D_L}{2\gamma}  \;.
\end{equation}
Notice that the current is proportional to the difference of diffusion coefficients. (In the quantum problem, to the difference $\bar{n}_1 - \bar{n}_L$.) On the contrary, the stationary actions $\tilde{I}_l$ (i.e., the diagonal elements of SPDM) are proportional to the sum of diffusion coefficients, 
\begin{equation}
\label{c5}
\tilde{I}_l=\frac{D_1+D_L}{2\gamma} \;.
\end{equation}
Exclusions are the first/last oscillator whose stationary actions is slightly above/below  than the value indicated  in Eq.~(\ref{c5}).
\begin{figure}
\includegraphics[width=12.5cm,clip]{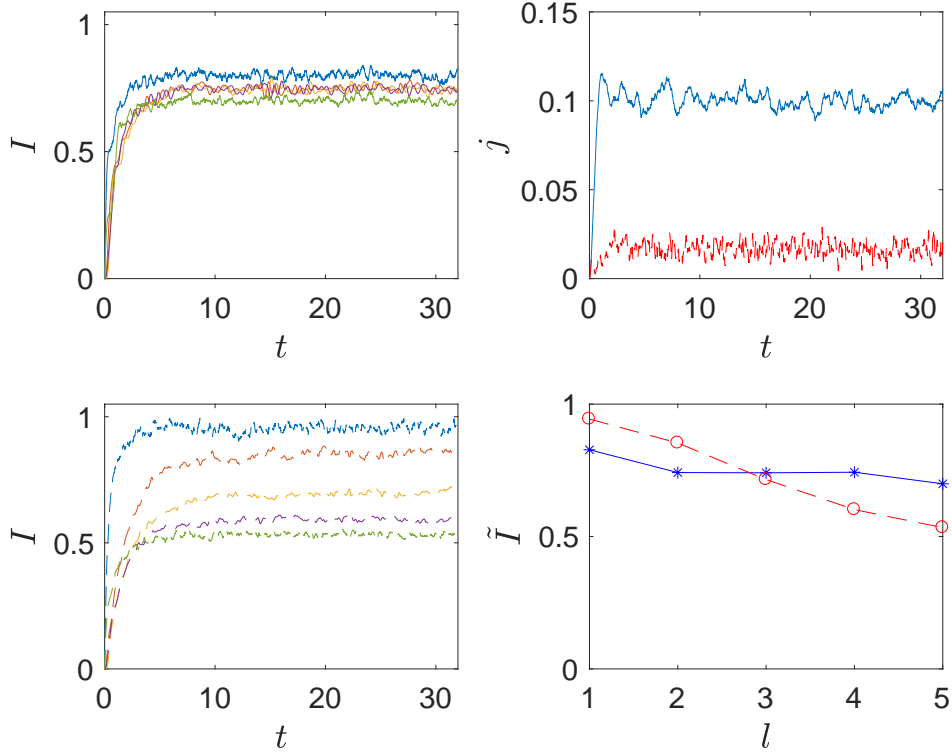}
\caption{Comparison between the cases $g=0$ and $g=2$ where the latter case is indicated by dashed lines. Dynamics of the mean actions $I_l(t)$, left column, dynamics of the mean current, upper-right panel, and stationary values of the actions $\tilde{I}_l$ along the chain, lower-right panel. The other parameters are $\gamma_1=\gamma_L=0.5$, $D_1=0.5$, and $D_L=0.25$. Average over 4000 runs.}
\label{fig2}
\end{figure}

\subsection{Nonlinear {\em vs.} linear chain}

We proceed with the case of interacting particles. To test our numerical code we first run it for $g=0$. The upper-left panel  in Fig.~\ref{fig2} shows dynamics of the diagonal elements $\rho_{l,l}(t)$ of SPDM for $L=5$, $\gamma_1=\gamma_L=0.5$, $D_1=0.5$, $D_L=0.25$, and the initial conditions $\rho_{l,l}(t=0)=0$, i.e., initially all oscillators are at rest. It is seen that oscillators rapidly approach the steady state with the stationary  actions $\tilde{I}_l$  given by Eq.~(\ref{c5}), see solid line in  lower-right panel. The dynamics of the mean current $j(t)$ is depicted in the upper-right panel by the solid blue line. Again, it is seen that the current approaches the value given by Eq.~(\ref{c4}).  The dashed lines in Fig.~\ref{fig2} depict the case of interacting particle.  A drastic decrease of the stationary current is noticed. 
\begin{figure}
\includegraphics[width=12.5cm,clip]{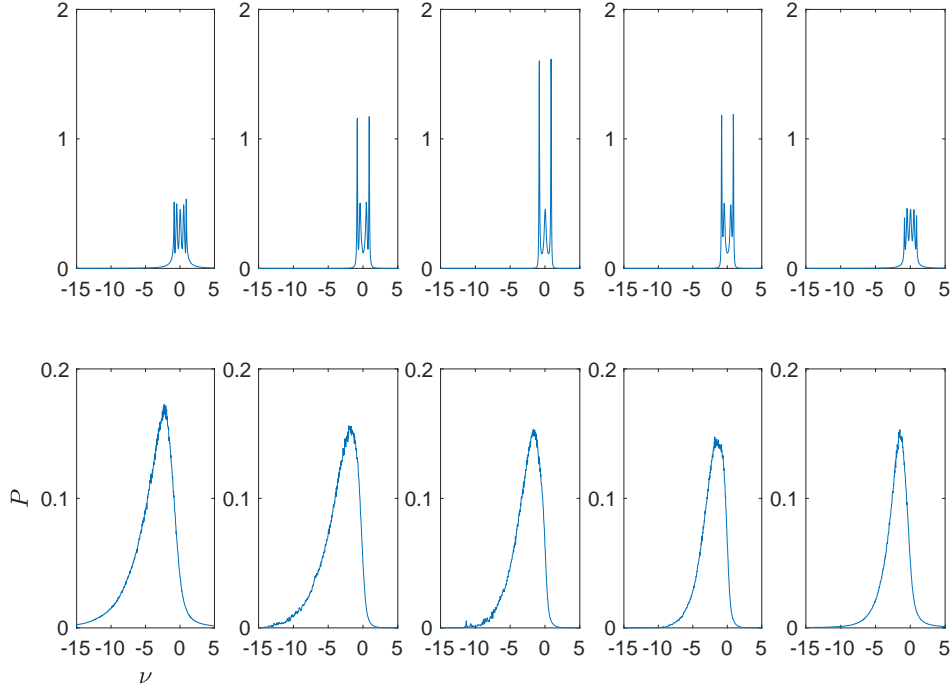}
\caption{Oscillators spectral density for $g=0$, upper row, and $g=2$, lower row. The other parameters are the same as in Fig.~\ref{fig2}.}
\label{fig3}
\end{figure} 
 
One finds a qualitative explanation for the observed effects by considering the oscillator spectral densities. Panels in the upper row in Fig.~\ref{fig3} show the spectral densities of oscillators in the chain for $g=0$.  It is easy to identify in this figure frequencies $\omega_k$ of the collective modes which are given by the real part of complex poles of the Green function for the system of linear differential equations (\ref{c2}) with omitted stochastic terms. For moderate values of the decay constants they can be approximated by eigenfrequencies of the linear chain,
\begin{equation}
\label{c7}
\omega_k=-J\cos(2\pi k/L)  \;,  \quad l=1,\ldots,L \;,
\end{equation}
which are located in the `conductance' band $|\omega_k|\le J$. Thus, the driving force acting on the first/last oscillator excites the collective modes. This leads to an efficient transport of excitations from the first to the last oscillator that quantum-mechanically corresponds to the flow of bosonic particles from the left to the right reservoir. Since the current is independent of the chain length $L$ we shall refer to this dynamical  regime as the ballistic transport. 
 
The lower panels in Fig.~\ref{fig3} show the spectral densities of oscillators for $g=2$. Now the system has no collective modes and transmission of excitations is a diffusion-like process, compare $P(\nu)$ in lower panels with $P(\nu)$ in the inset in Fig.~\ref{fig1}. As a consequence, we observe considerable decrease in the stationary current. Furthermore, contrary to the ballistic regime, in the diffusive regime the current depends on the chain length $L$.  We discuss this dependence in the next subsection.
\begin{figure}
\includegraphics[width=12.5cm,clip]{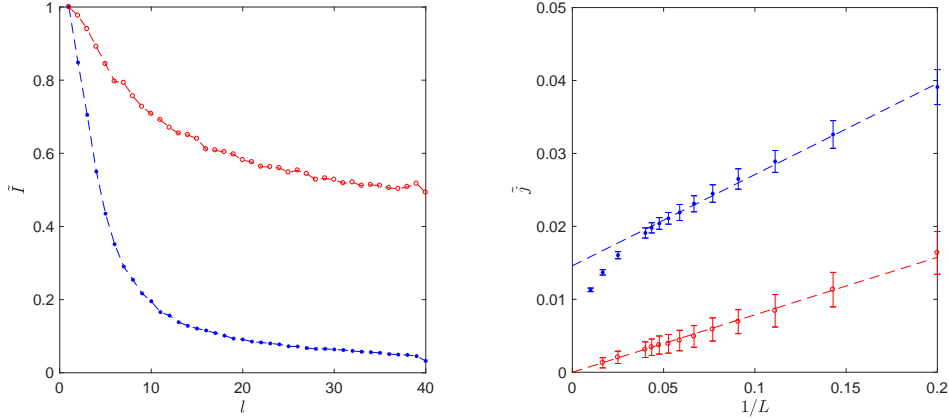}
\caption{Right panel: The stationary current as the function of the inverse chain length for $g=2$, $\gamma_1=\gamma_L=0.5$, $D_1=0.5$, and $D_L=0.25$ (open circles) and $D_L=0$ (asterisks). The bars show statistical error due to finite number of realizations (4000). Left panel: The mean oscillator actions along the chain of length $L=40$ for two considered cases $D_L=D_1/2$ (open circles) and $D_L=0$ (asterisks).}
\label{fig4}
\end{figure} 

\subsection{The limit of long chains}

We performed extensive numerical simulations of the system dynamics for a large system size up to $L=100$.  The open circles in Fig.~\ref{fig4}(b)  show the stationary current $\tilde{j}$ as the function  of $1/L$. The linear dependence is noticed that is consistent with the above conjecture about the diffusive transport. We mention that the change of the transport regime from ballistic to diffusive with increase of $g$ is a crossover rather than a phase transition.  The necessary condition for this crossover is that the nonlinear shift of the oscillator frequencies, which is estimated as $g\tilde{I}_l$, exceeds the width of the conductance band, i.e.,  
\begin{equation}
\label{c8}
\tilde{I}_l > J/g  \;.
\end{equation}
We stress that the condition (\ref{c8}) has to be satisfied for {\em all} oscillators to get the universal  dependence $\tilde{j}\sim 1/L$. This, for example, is not the case if we set $D_L$ to zero. (In the original quantum problem this corresponds to pure sink of particle at the right edge of the chain.)  Here the condition (\ref{c8}) is satisfied for oscillators in the left part of the chain but violated for oscillators in the right part. As the consequence,  the spectral densities of oscillators in the left and right parts of the chain resemble those for the diffusive and ballistic transport, respectively. The dependence of the stationary current on the system size for this mixed transport regime in depicted by asterisks in Fig.~\ref{fig4}(b). It is seen that the presence of  `ballistic part'  results in a larger current as compared to the pure diffusive transport regime. The depicted numerical data also indicate highly nontrivial dependence of the stationary current on the difference $\bar{n}_1-\bar{n}_L$ which involves the chain length $L$ as an additional parameter.

\section{Conclusion}

We analyze stationary current of the bosonic carriers in the Bose-Hubbard chain of length $L$ where the first and the last sites of the chain are attached to reservoirs of Bose particles. The analysis is carried out by using the pseudoclassical approach which reduces the original quantum problem to the classical problem for $L$ coupled oscillators where the first and last oscillators are subject to both friction and excitation. In the case of vanishing inter-particle interactions we analytically calculate elements $\rho_{l,m}$ of the single-particle density matrix  which classically correspond to the correlation functions between the $l$th and $m$th oscillators. It is shown that the pseudoclassical approach reproduces the quantum-mechanical results {\em exactly}. In particular, the stationary current (which is determined by off-diagonal elements of the stationary single-particle density matrix) is proportional to the particle density difference $\Delta \bar{n}=\bar{n}_1-\bar{n}_L$ of the reservoirs and is independent of the chain length $L$.

Next we numerically address the case of finite inter-particle interactions which make the classical oscillators nonlinear. A drastic reduction of the stationary current as compared to the linear case is noticed. Furthermore, the current depends on the chain length and in the case of small difference  $\Delta \bar{n} \ll  \bar{n}_1$ scales as $1/L$. We give a qualitative explanation for this change in the transport regime by introducing and analyzing the spectral densities of oscillators. We mention that the oscillator spectral density explicitly involves the notion of trajectory and, hence, is meaningful only in the classical approach. Nevertheless, it allows us to explain peculiarities of the quantum transport in the whole parameter space of the system.

\acknowledgments
This work has been supported through Russian Science Foundation grant N19-12-00167.


\end{document}